# The Ultrafast Superconducting Diode Effect

E. Wang[1†], M. Chavez-Cervantes[1], J. Satapathy[1], T. Matsuyama[1], G. Meier[1], X. Zhang[2], L. You[2], F. Marijanovic[3], J.B. Curtis[3], E. Demler[3], A. Cavalleri[1,4†]

*1 Max Planck Institute for the Structure and Dynamics of Matter, Hamburg, Germany*

*2 Shanghai Key Laboratory of Superconductor Integrated Circuit Technology, Shanghai Institute of Microsystem and Information Technology, Chinese Academy of Sciences, 200050 Shanghai, China*

*3 Institute for Theoretical Physics, ETH Zurich, 8093 Zurich, Switzerland*

*4 Clarendon Laboratory, University of Oxford, Parks Road, Oxford OX1 3PU, United Kingdom*

† Corresponding authors: eryin.wang@mpsd.mpg.de; andrea.cavalleri@mpsd.mpg.de

**Nonreciprocal transport is generally observed in superconductors in which time reversal and inversion symmetries are simultaneously broken. This effect, which may become one of the backbones for future superconducting electronics, arises because of asymmetric vortex transport in a magnetic field. However, vortex transport is also intrinsically dissipative and limited in speed. Here, we report on the discovery of ultrafast non-reciprocal transport in centrosymmetric superconductors. For NbN films biased with a quasi-DC supercurrent, picosecond current pulses with the same sign as the bias experience resistive impedance, whereas pulses of opposite polarity encounter an inductive response. Strikingly, the effect is at least three orders of magnitude faster than in conventional superconducting diodes, limited only by ultrafast current-induced depairing. We demonstrate rectification of a 100 GHz signal, with dissipation levels of a few fJ per cycle. We foresee potential for superconducting logic elements, operating at THz bit rates with aJ energy dissipation per operation.**

Ultrafast processes in superconductors are of fundamental interest and are relevant for energy-efficient electronic technologies. A key challenge for practical super-conducting electronics is the realization of functional analogs of semiconductor logic elements. Amongst these, the superconducting diode has attracted strong interest [1-15]. One established route to diode functionality relies on controlling vortex entry and motion in asymmetric superconducting structures [2-6, 12-15]. However, vortex-based diodes (Fig. 1a) are inherently limited by the slow dynamics of vortices, which restrict the operation speed. To date, the fastest reported vortex diode operation reaches ~100 MHz [15].

The speed limit of the conventional superconducting diode effect is underscored by the experiments reported in figure 1, in which we compared the response of a NbN strip to microsecond and picosecond current pulses. A 20-nm-thick, 8-μm-wide NbN strip ($T_c$ = 14.5 K) patterned with triangular notches along one edge (Fig. 1b; fabrication details in Supplementary Information) was used for this purpose. Unless otherwise noted, all measurements shown in this paper were performed at $T$ = 7 K.

In the absence of an external magnetic field, the critical current densities $J_c$ for the two current polarities ($+I_{DC}$ and $-I_{DC}$) were measured to be identical, with $J_c \approx 100$ GA/m$^2$, defined by the narrowest section of the strip (width ≈ 6 μm). On the other hand, when a perpendicular magnetic field $B_{ext}$ was applied and aligned with the current-induced field $B_{\pm I_{DC}}$ at the notched edge, non-reciprocal transport emerged. As extensively documented in the literature, this diode effect is rooted in vortex dynamics at the notches, where the local magnetic field is modified, either facilitating or suppressing vortex penetration and motion, depending on the current polarity. As a result, $+I_{DC}$ and $-I_{DC}$ exhibited different critical current densities. Reversing the direction of $B_{ext}$ correspondingly reversed the diode polarity (Fig. 1b).

This vortex-based diode effect was not observed for picosecond current pulses. We integrated the notched NbN strip into a picosecond transport platform [16-22] (Fig. 1c), connecting the NbN strip to photoconductive switches via a coplanar waveguide (see Supplementary Information). We define the transmittance as the ratio between the peak current of the pulse transmitted through the sample and that of the incident pulse. Because the transmittance scales with the superfluid density (see Supplementary Information; see also Ref.[23]), 1–Transmittance can be used as a measure of superconductivity suppression under increasing peak current density $J_{\mathrm{peak}}$ of picosecond current pulses ($\pm I_{\mathrm{ps}}$).

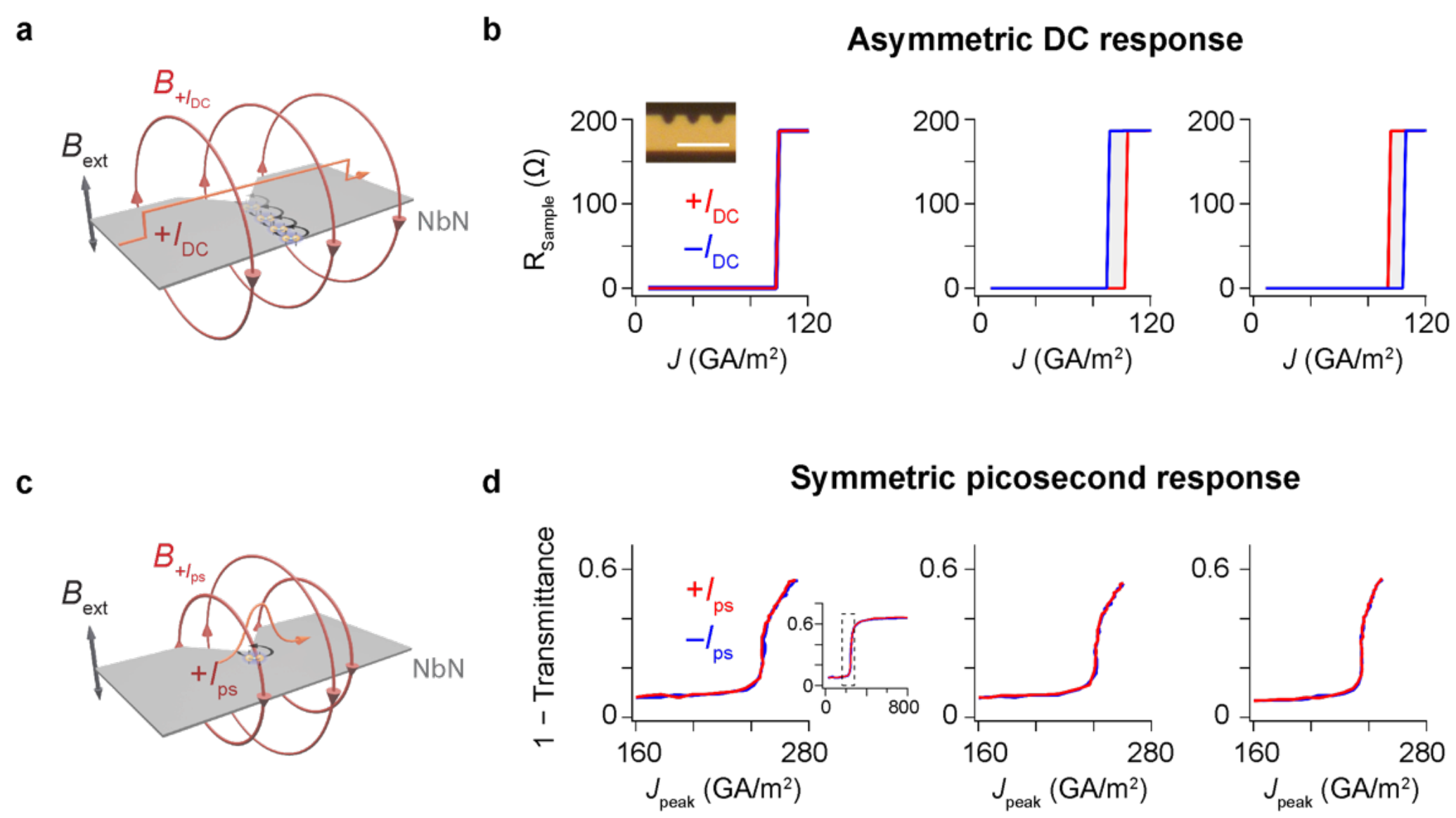

**Figure 1 | Quasi-DC and picosecond transport measurements of the superconducting vortex diode effect.** (a) Schematic of field distribution in quasi-DC measurements on a NbN strip patterned with triangular notches along one edge, under a perpendicular external magnetic field $B_{\mathrm{ext}}$. Vortex penetration occurs at a lower current when $B_{\mathrm{ext}}$ aligns with the current-induced field $B_{\pm I_{DC}}$ at the notched side, enhancing the local magnetic field. (b) Critical current densities for both current polarities measured at $B_{\mathrm{ext}}$ = 0 (left), downward $B_{\mathrm{ext}}$ (middle), and upward $B_{\mathrm{ext}}$ (right). Inset: optical micrograph of the notched NbN strip; scale bar, 10 μm. (c) Schematic of field distribution in picosecond (ps) transport measurement. (d) 1-Transmittance as a function of peak current density for $B_{\mathrm{ext}}$ =0 (left), downward (middle) and upward (right). 1-Transmittance quantifies the suppression of superconductivity (see text). Inset: enlarged view for the zero-field data. All measurements were performed at $T$ = 7 K below $T_c$ = 14.5 K. The magnitude of $B_{\mathrm{ext}}$ is ~ 2.6 mT. The slight reduction in the depairing threshold is attributed to the weak suppression of superconductivity by the magnetic field. These results establish that ultrafast superconducting diode operation cannot rely on vortex dynamics and instead requires a fundamentally different mechanism.

At zero magnetic field, we observed suppressed superconductivity for picosecond current pulses at $J_{peak} \approx 250$ GA/m$^2$ (Fig. 1d). Note that this value is far larger than the dc critical current density $J_c$, and was interpreted in previous experiments [23] as an indication of current induced depairing. This phenomenon is limited only by the local response of the superconducting condensate and occurs at timescales over which vortices are frozen in place, enabling access to higher peak currents than for longer current pulse durations. Unlike the slow dynamics reported in figure 1a, even under an external magnetic field, the picosecond responses to $+I_{ps}$ and $-I_{ps}$ remain symmetric (Fig. 1d), consistent with the notion that vortices are inertially frozen at picosecond timescales.

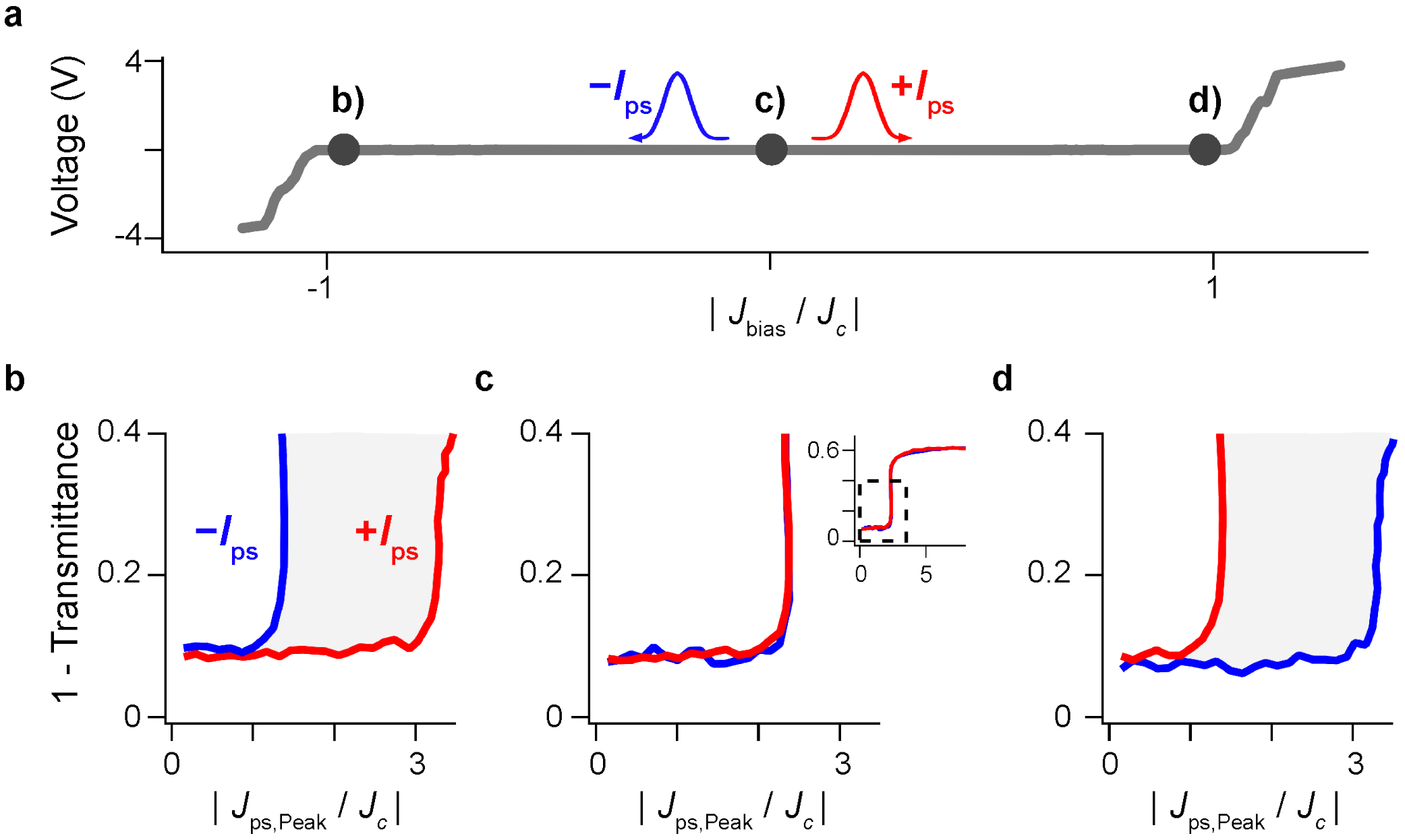

**Figure 2 | Bias-current modulation of the picosecond response.** (a) Sample voltage as function of applied quasi-DC current density. Black dots denote the pre-pulse state prior to the application of picosecond current pulses ($\pm I_{ps}$) in panels (b-d). (b-d) 1-Transmittance as a function of normalized transmitted peak current density under negative (left), zero (middle) and positive bias current $J_{bias}$ ($|\pm J_{bias}| = 0.97\ J_c$). Inset: Enlarged view of the zero-bias data.

As sketched in figure 2a, the transport reciprocity observed at picosecond timescales is expected to be broken under a quasi-DC current bias. Intuitively, one expects that when a current pulse of the same polarity as the bias traverses the superconductor, the response can be made resistive by exceeding the depairing limit. On the contrary, for the same current bias, a pulse of the same amplitude but opposite polarity should experience superconductive response.

In figure 2 we report experiments performed in a notch-free NbN strip (20 nm thick, 10 μm wide and 30 μm long), integrated into a coplanar waveguide. The bias current was applied by an external nanosecond pulse generator (see Supplementary Information). The current density of the nanosecond pulse was set to $0.97J_c$, ensuring that superconductivity was not suppressed by the bias alone. As expected, a positive bias current $+I_{bias}$, lowers the depairing threshold for $+I_{ps}$ while raising it for $-I_{ps}$. Reversing the polarity of $I_{bias}$ correspondingly reverses the asymmetry.

The effect discussed above was shown to operate at least at 100 GHz frequency. To demonstrate the high frequency response beyond the single pulse operation of figure 2, we studied the response to a periodic train of bipolar pulses, driving the current in both directions starting from the bias level set by the quasi-DC current. The experimental architecture is shown in Fig. 3a. Bipolar current pulse trains were generated by illuminating the upper and lower photoconductive switches — biased at $+V$ and $-V$, respectively — with two interdigitated pairs of double laser pulses, driving four pulses of alternating polarity separated by 5-ps time delay (see Supplementary Information). The pulse amplitude, sampled by an adjacent switch to the left of the launch point, exceeded the depairing threshold (Fig. 3b), resulting in a reduced transmission through the sample (Fig. 4c). In the absence of a bias current, the device response was found to be symmetric, producing no measurable dc offset. On the other hand, the application of a nanosecond bias current $I_{bias}$ (~ 0.97$J_c$) induced the asymmetric picosecond response shown in Fig. 2 and led to rectification (Fig. 3 d,e). The small asymmetry between rectified signals for +$I_{bias}$ and −$I_{bias}$ (Fig. 3d, e) likely arises from slight variations in the amplitudes of the bipolar pulses, which are amplified by the strong nonlinearity near the sudden drop of transmittance (Fig. 2c). Note that the polarity of the rectified signal is directly controlled by the sign of $I_{bias}$.

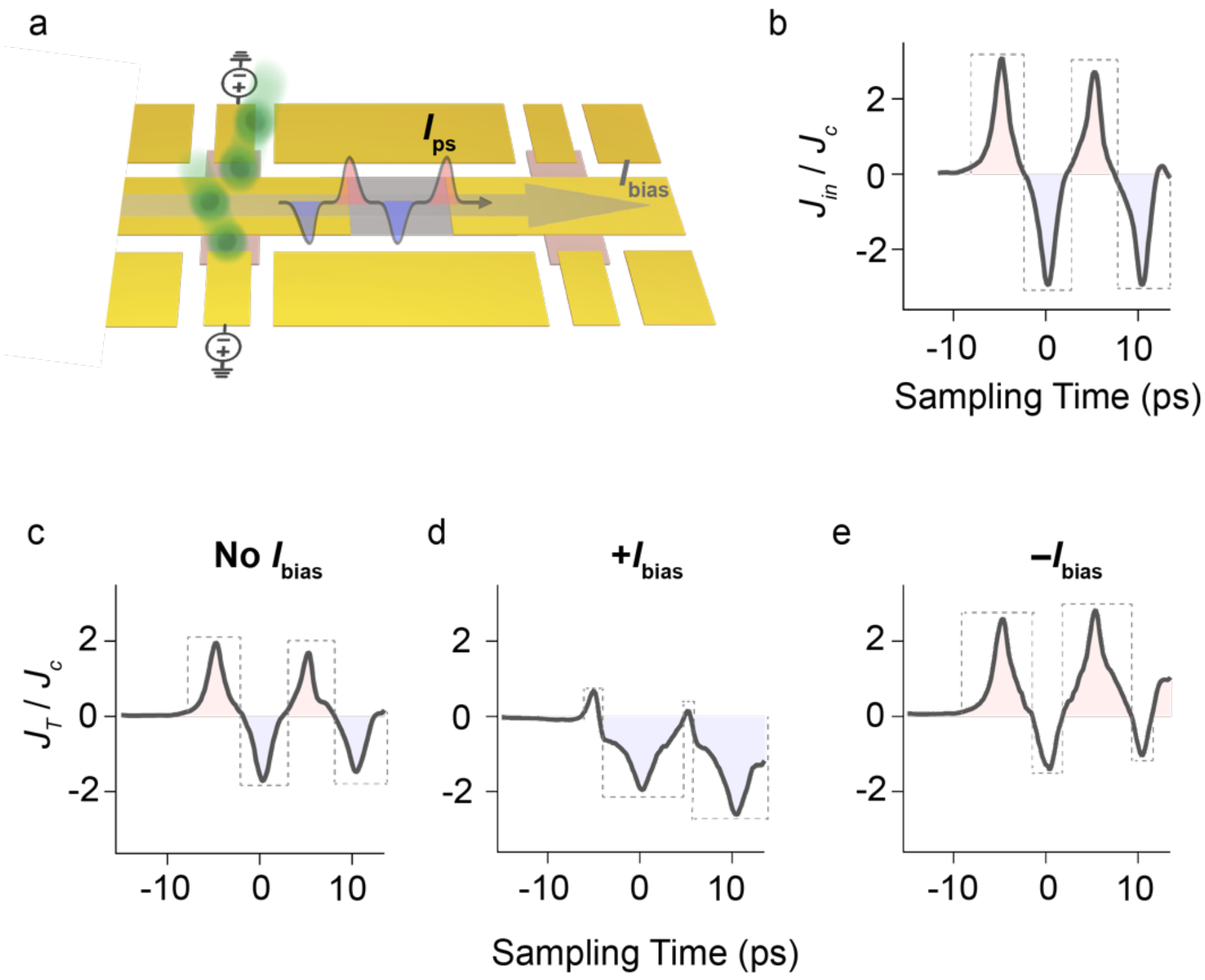

**Figure 3 | Rectification of bipolar picosecond current pulses by a bias current $I_{bias}$.** (a) Experimental schematic. The left pair of photoconductive switches (grey) are biased independently with positive and negative voltages. Two sets of laser pulses (each consisting of dual pulses separated by 10 ps) independently trigger the two switches, generating bipolar current pulses. The launched pulses propagate through the sample and are detected by one of the photoconductive switches on the right. (b) Time profile of the launched bipolar current pulse, measured using an adjacent photoconductive switch located to the left of the launch switch. (c-e) Time profiles of transmitted current pulses for zero (left), positive (middle) and negative (right) $I_{bias}$.

The identical responses of picosecond current pulses with opposite polarities under an external magnetic field — shown in Fig. 1b — exclude vortex dynamics as the origin of the observed picosecond current threshold, above which superconductivity is abruptly suppressed. Instead, this behavior is attributed to Cooper pair depairing, caused by the current-induced Doppler energy shift of the Bogoliubov quasiparticle spectrum, $\hbar k_F \cdot v_s$, becoming comparable to the superconducting gap $\Delta$ (where $k_F$ is the Fermi momentum and $v_s$ is the velocity of Cooper pairs) [23, 24]. Consequently, the quasiparticle bands shift downward for positive wave vectors and upward for negative wave vectors (Fig. 4a). Upon application of a bias current $I_{\text{bias}}$, the quasiparticle spectrum is pre-shifted, leading to asymmetric depairing thresholds for subsequently applied positive and negative picosecond current pulses ($\pm I_{\text{ps}}$): the depairing threshold is reduced for $+I_{\text{ps}}$ and enhanced for $-I_{\text{ps}}$.

The depairing dynamics can be modeled within microscopic BCS theory in the dirty limit using the Usadel equation [25], with material parameters extracted from experimental characterizations of the sample [23]. The sharp transmission drop is attributed to the strong-nonlinearity of the superfluid density near the depairing threshold, which results from the isotropic s-wave gap size and strong disorder of the system. By incorporating the quasiparticle spectrum pre-shift induced by $+I_{\text{bias}}$, the calculations reproduce the asymmetric picosecond depairing thresholds, as shown in Fig.4b (see supplementary information for details). This agreement confirms that the asymmetric picosecond responses originate from ultrafast modulation of the Bogoliubov quasiparticle spectrum rather than vortex dynamics.

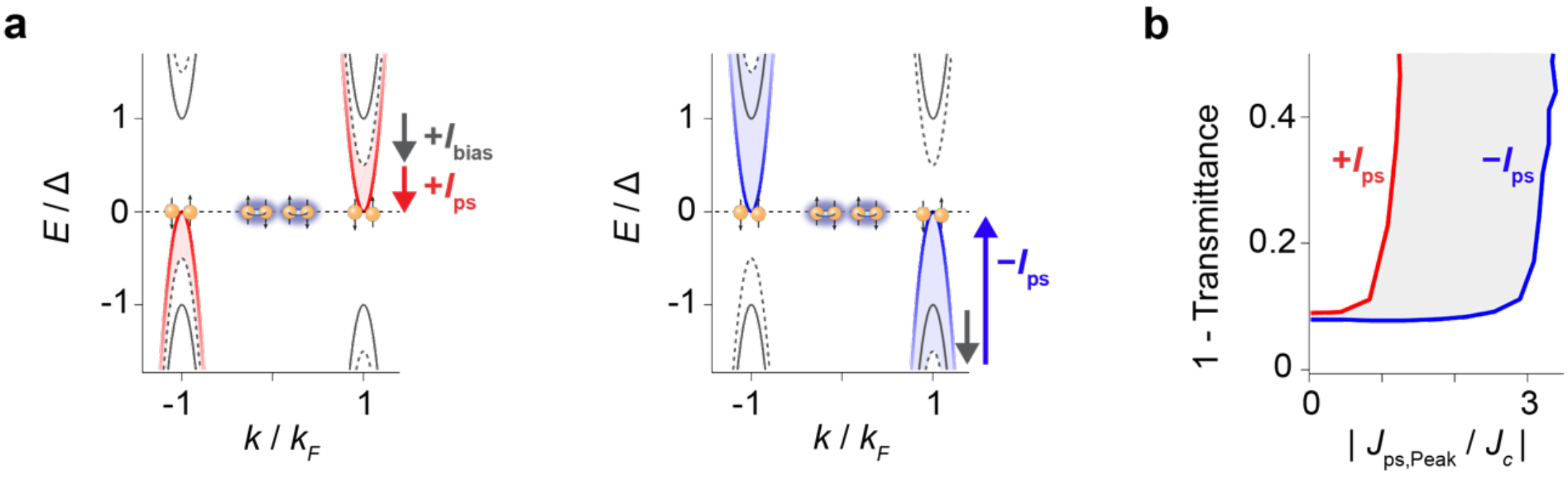

**Figure 4| Bias-current-induced asymmetric depairing by +$I_{ps}$ and −$I_{ps}$.** (a) Left: Bogoliubov quasiparticle spectrum shifts under the combined application of a bias current +$I_{bias}$ and +$I_{ps}$. The supercurrent shifts the quasiparticle spectrum by $\Delta E = \hbar k_F \cdot v_s$, where $k_F$ is the Fermi momentum and $v_s$ is the Cooper pair velocity. Under a positive current, the quasiparticle bands shift downward for positive wave vectors and upward for negative wave vectors. The equilibrium quasiparticle bands are shown as solid grey lines. The bands shifted by +$I_{bias}$ alone are indicated by dashed grey line while the bands after further application of +$I_{ps}$ are shown by solid red lines. Right: same as left, but with the application of a negative picosecond current −$I_{ps}$. The quasiparticle bands after further application of −$I_{ps}$ are shown as solid blue lines. (b) Simulated asymmetric depairing thresholds for positive and negative picosecond current pulses with application of +$I_{bias}$ ~ 0.97 $J_c$.

The depairing dynamics demonstrated here bypasses slow vortex motion and provides a new route for designing ultrafast, field-free superconducting electronics. The fundamental speed limit of this mechanism is set by the relaxation time of the superconducting order parameter, which is not known precisely, but is expected to be near or below 1 ps, as evidenced by our multi-pulse measurements. Future work will characterize this effect in all-electrical pump-probe experiments. However, we foresee that logic elements [26] based on the superconducting diode effect demonstrated here could operate at terahertz frequencies, with an estimated energy dissipation per operation on the order of a few fJ with a device like the one used for the experiments reported in figure 3. This dissipation can be further reduced to the aJ level through straightforward device miniaturization (see Supplementary Information for details).

In summary, we have demonstrated a picosecond superconducting diode effect based on ultrafast depairing. By circumventing vortex dynamics, this approach overcomes the fundamental speed limitations of vortex-mediated superconductor diodes and enables rectification at frequencies near 100 GHz, which could likely be extended to the THz range. These operation speeds are already three-to-four orders of magnitude faster than conventional vortex diodes achieved so far. More broadly, nonlinear current transport in superconductors may become applicable to other fields of science and technology, ranging from sensing in SQUID magnetometers and potentially to quantum electrodynamic circuit applications.

# Supplementary information for
# The Ultrafast Superconducting Diode Effect

## S1. Device fabrication

NbN film samples were epitaxially grown on MgO substrates by DC magnetron sputtering, with the substrate temperature maintained at 300 ℃. The films had a thickness of approximately 20 nm and exhibited a superconducting transition temperature of $T_c \approx 14.5$ K [1]. NbN thin films were patterned using laser lithography followed by reactive ion etching (RIE) with $SF_6$.

Photoconductive switches and coplanar waveguides were fabricated on MgO substrates with pre-patterned NbN thin films using laser lithography and electron-beam deposition. Specifically, a 200 nm-thick amorphous silicon layer was deposited to form the photoconductive switches, while 10 nm Ti / 260 nm Au was deposited to define the coplanar waveguides. To ensure good electrical contact between the waveguides and the NbN films, the NbN contact regions were cleaned by argon plasma and subsequently capped *in situ* with 50 nm Au layer prior to Ti/Au deposition. Using this procedure, a contact resistance of ≤ 3 Ω was achieved for each contact.

## S2. Device characterization

In total, three devices were used in this study, hereafter referred to as device #1, #2 and #3. Figure S1 shows an optical micrograph of device #2, from which the data presented in Fig. 2 were obtained. The dimensions of the NbN strip are approximately 10 μm in width, 30 μm in length, and 20 nm in thickness.

Device #1 and #3 share a similar overall architecture with device #2, differing only in specific design details. The dimensions of the NbN strips in device #1 and #3 are approximately 8 μm in width, 30 μm in length, and 20 nm in thickness. Device #3, used for the bipolar pulse-train experiments shown in Fig. 4, features a signal line width at the photoconductive switch positions of approximately 40 μm—twice that of device #2—to enable efficient generation of bipolar pulses, as discussed in Section S.8. Device #1 is identical to device #3, except that three triangular notches were patterned along one edge of the NbN strip, as shown in Fig. 1. The data presented in Fig. 1 were obtained from device #1.

The characterization of the sample's superconducting coherence length and magnetic penetration depth are shown in a previous paper [2].

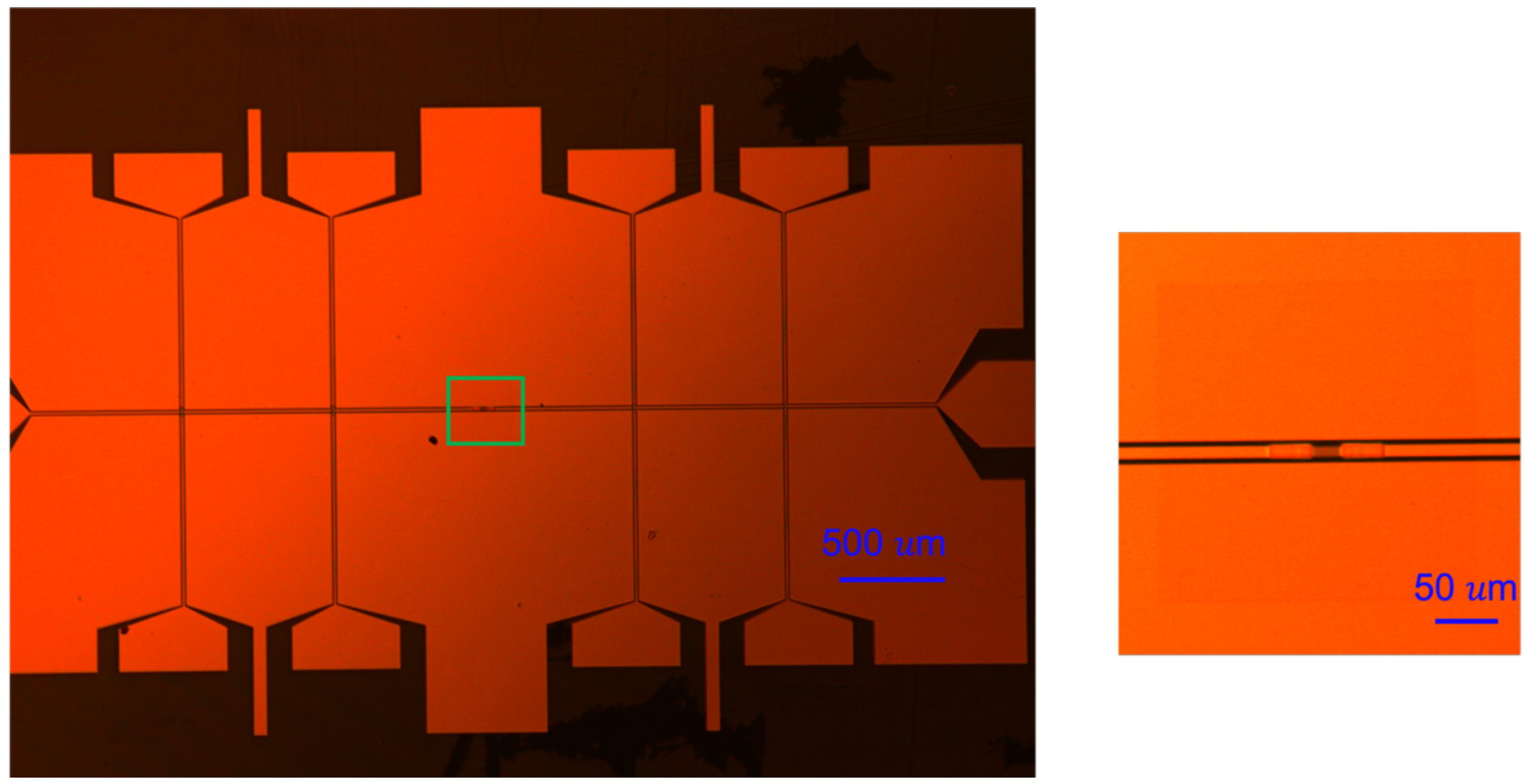

**Figure S1| Optical micrograph of NbN device #2.**

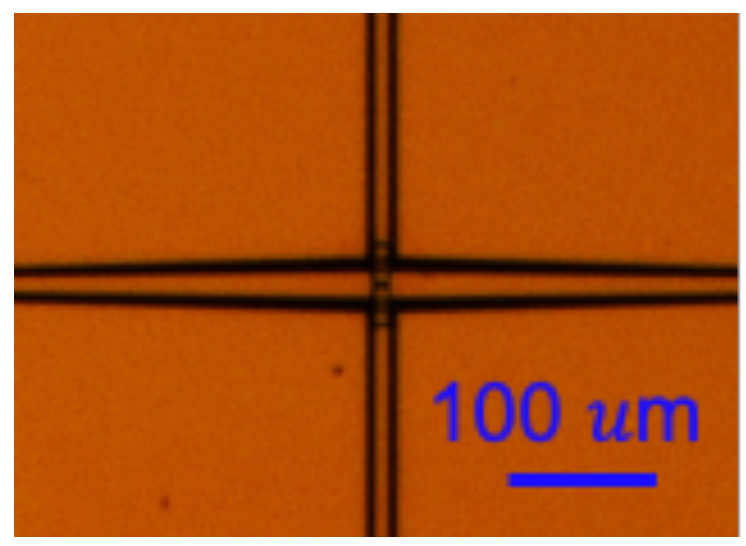

**Figure S2| Optical micrograph of the widened coplanar waveguide region near the switches in device #3.**

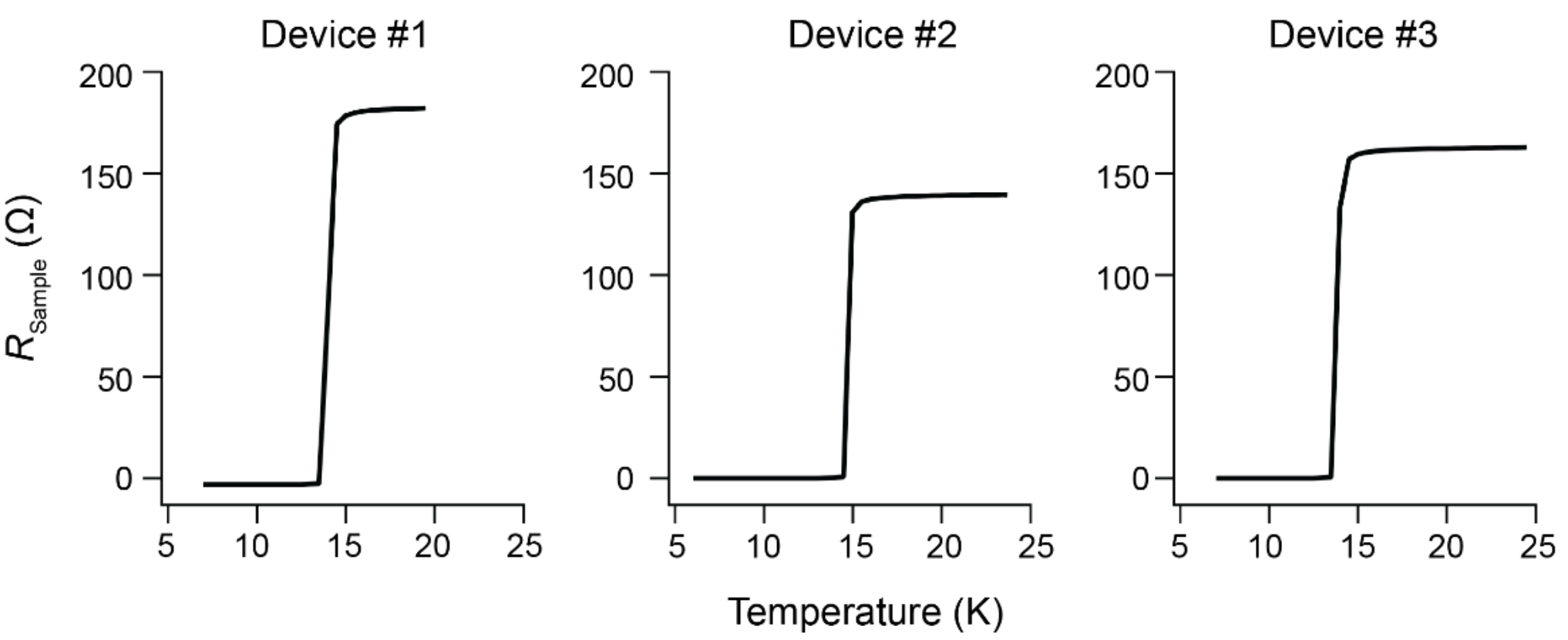

**Figure S3| Superconducting transition in resistance versus temperature measurements for device #1, #2 and #3.** The bias current density is 5 $MA/m^2$.

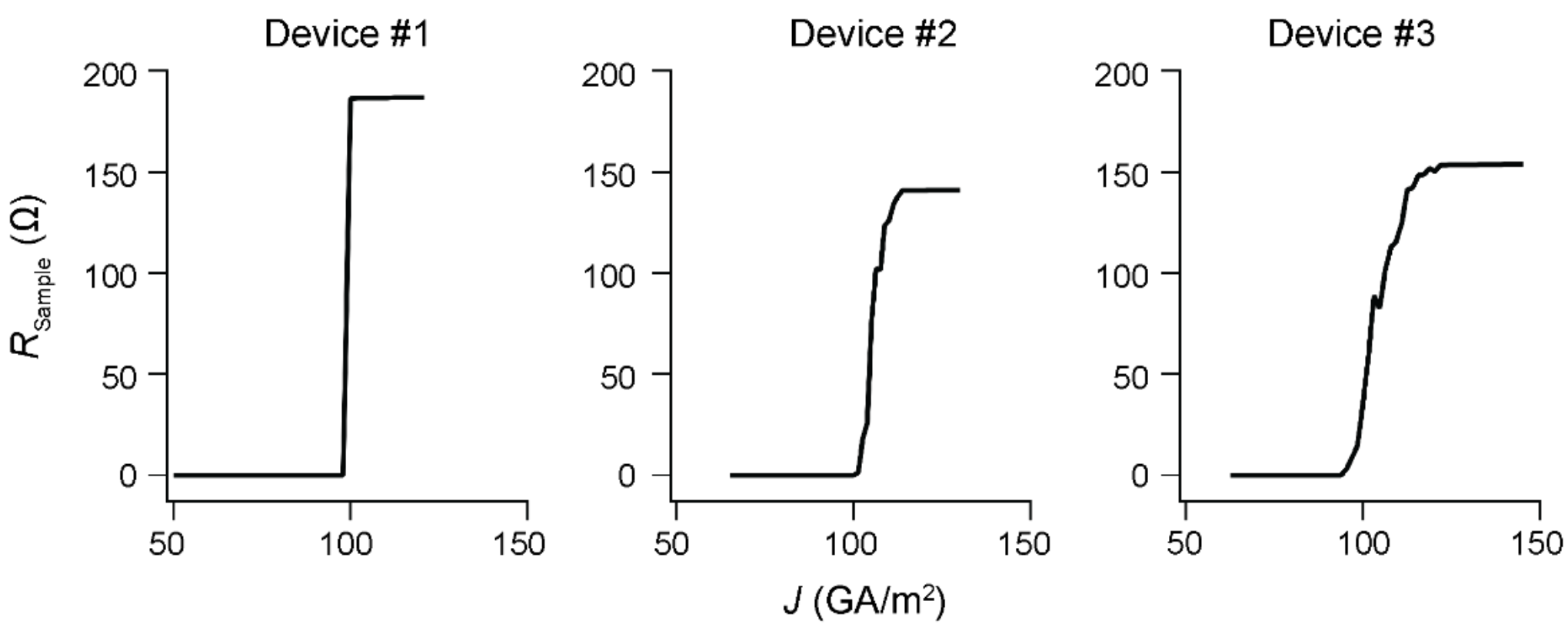

**Figure S4| Critical current measurements for device #1, #2 and #3 at $T$ = 7 K.**

## S3. Measurement geometry

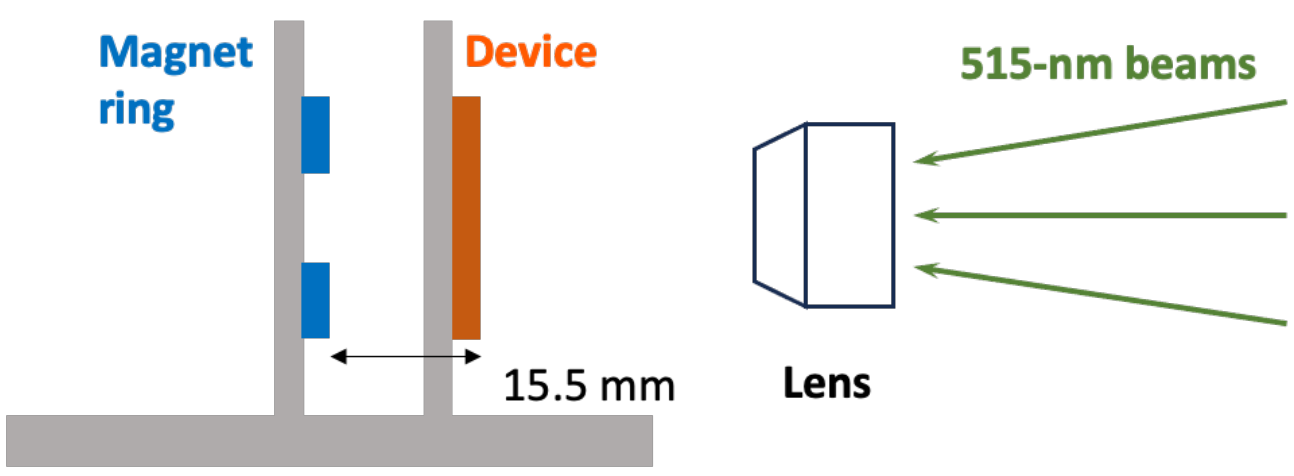

**Figure S5| Measurement geometry.**

The magnetic field was applied using a permanent ring magnet (Nd-Fe-B, grade N45; inner/outer diameter ~ 9/12 mm; thickness ~ 1.5 mm), positioned approximately 15.5 mm directly behind the device, as shown in Fig. S5. The 515-nm laser beams were focused onto the device from the front side using a lens. Based on the magnetization–temperature dependence reported in Ref. [3], we estimate that the magnetic field generated by the ring at the position of the NbN strip is ~ 2.6 mT.

The direction of the magnetic field was reversed by manually flipping the magnet ring after warming up the cryostat. Quasi-DC and picosecond transport measurements under a given magnetic-field configuration were performed consecutively without disturbing the magnet position. Owing to slight changes in the magnet alignment after reversal, the critical currents and depairing currents measured under downward and upward magnetic fields shown in Fig. 1 exhibit small quantitative differences.

## S4. Picosecond transport measurement

The 515-nm, 250-fs-long laser beams — used for exciting the photoconductive switches — were seeded from a pharos laser operating at 50 kHz repetition rate. The laser beam used to launch the picosecond current pulses was modulated by a mechanical chopper at 1 kHz. The total charge collected from the photoconductive switches used to probe the current pulses was amplified using custom-built ultralow-noise transimpedance amplifiers and subsequently measured with a lock-in amplifier.

The calibration procedures follow those described in Refs. [4, 5]. Briefly, the sensitivity of each photoconductive switch used for probing current pulses was calibrated using the configuration shown in Fig. S6a. In this setup, the signal line was biased with a DC voltage source while the opposite end was left open. The switch under calibration was illuminated with a 515-nm laser beam chopped at 1 kHz, and the resulting signal was measured using the transimpedance amplifier and lock-in amplifier. A linear dependence of the measured signal on the applied bias voltage was observed. As representative examples, calibration results for the switches measuring the incident and transmitted current pulses at 7 K in device #2 are shown in Figs. S6b and S6c, respectively. All quantities measured using photo-excited switches were normalized by the slope α of the corresponding calibration curve.

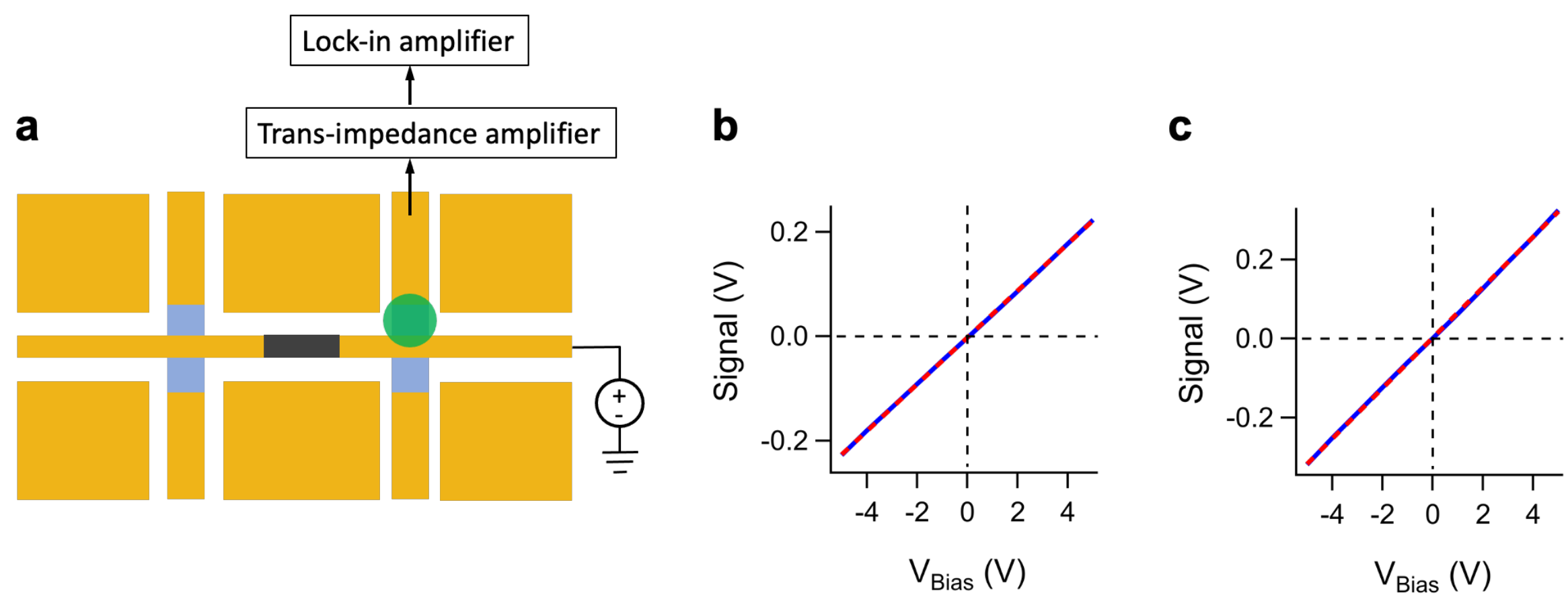

**Figure S6| Switch calibration.** (a) Sketch for calibration process. (b) Calibration of the photoconductive switch used for detecting incident pulses. The measured curve is shown in blue and the linear fitting in dashed red. (c) Same as (b), but for the photoconductive switch used for detecting the transmitted pulses.

The peak current amplitudes of the incident and transmitted pulses were calculated using

$$I_{peak} = \frac{1.414\, V_{signal,peak}}{\alpha \cdot Z_0 \cdot 0.85} \qquad \text{(S1)}$$

where $V_{signal,\, peak}$ is the measured peak signal from the amplifier, α is the slope of the photoconductive switch calibration curve discussed above, and $Z_0 \approx 50\ \Omega$ is the characteristic impedance of the coplanar guide. The factor 0.85 accounts for the average transmission between the pulses incident on the photoconductive switch and those transmitted through it, taking into consideration an approximately 30% reflection caused by local impedance mismatch (see Ref. [4]), yielding an average factor of (1+0.7)/2 = 0.85. The prefactor 1.414 accounts for the correlation between the actual current pulse temporal profile and the response function of the photoconductive switch. This approximation is justified because the two profiles are expected to be very similar, given the high bandwidth of the coplanar waveguide (~0.7 THz at −3 dB) compared with the full width at half maximum of the current pulses (FWHM ≈ 2 ps). This calibration method has been validated previously (see Ref. [2]).

The temperature dependence of the transmitted pulse measured on device #2 is shown in Fig. S7. The dependence of the transmitted pulse peak current amplitude at $T = 7$ K ($T < T_c$) for device #2 is shown in Fig. S8. As can be seen, the Transmittance (defined as the ratio between the peak current of the pulse transmitted through the sample and that of the incident pulse) decreases with lower superfluid density, thus 1-Transmittance can serve as a measure of suppression of superconductivity.

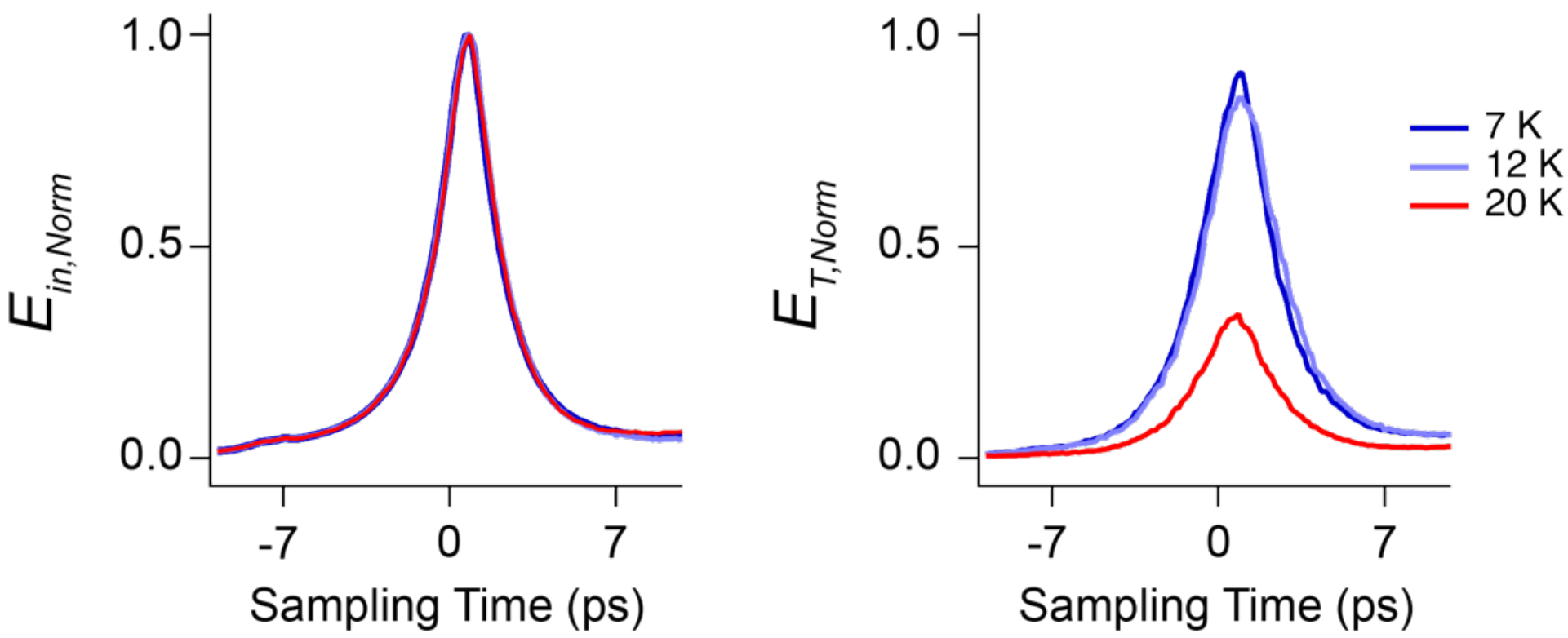

**Figure S7| Temperature dependence of picosecond measurement.** Normalized electric field time profiles of incident (left) and transmitted (right) current pulses at $T = 7$, 12 and 20 K. The electric field is normalized by the peak field of the incident current pulse.

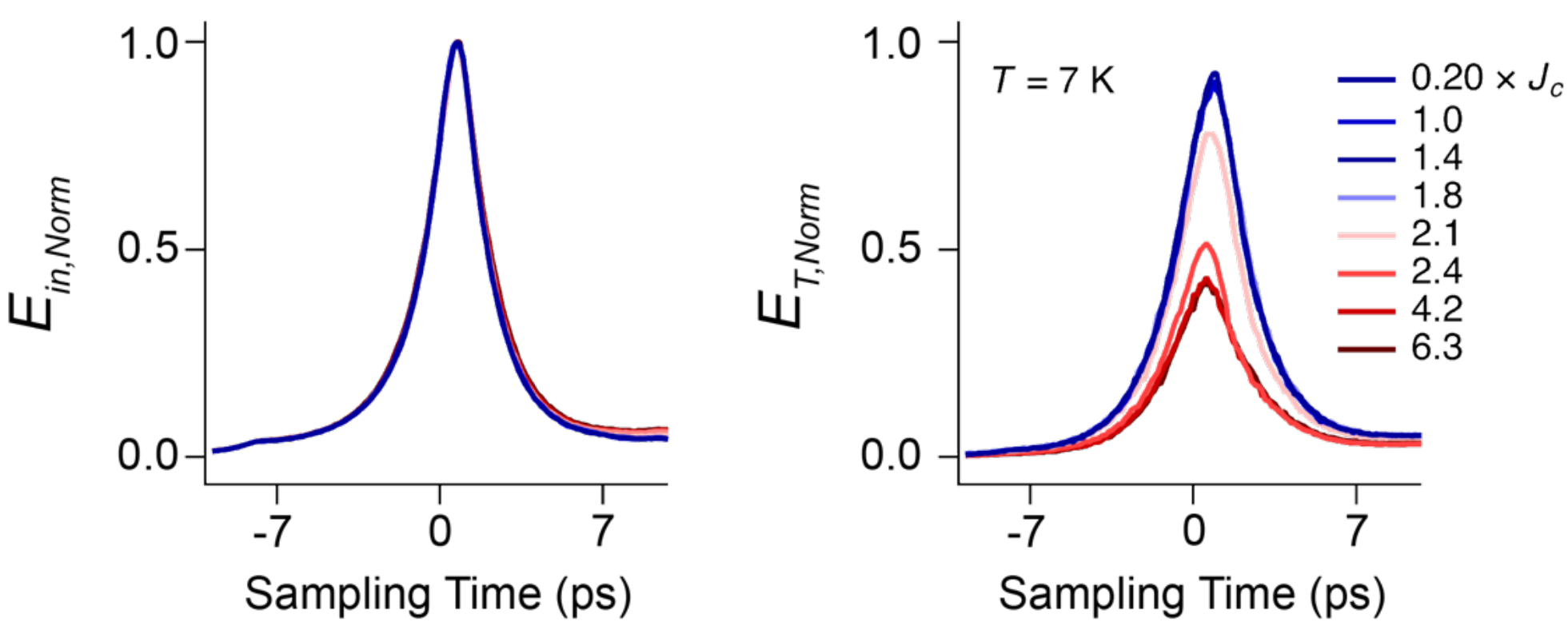

**Figure S8| Peak current amplitude dependence of picosecond measurement.** Normalized electric field time profiles of incident (left) and transmitted (right) current pulses for different transmitted peak current densities at $T = 7$ K. The electric field is normalized by the peak field of incident current pulse.

## S5. Bogliubov quasiparticle spectrum shift by $\pm I_{bias}$ and $\pm I_{ps}$

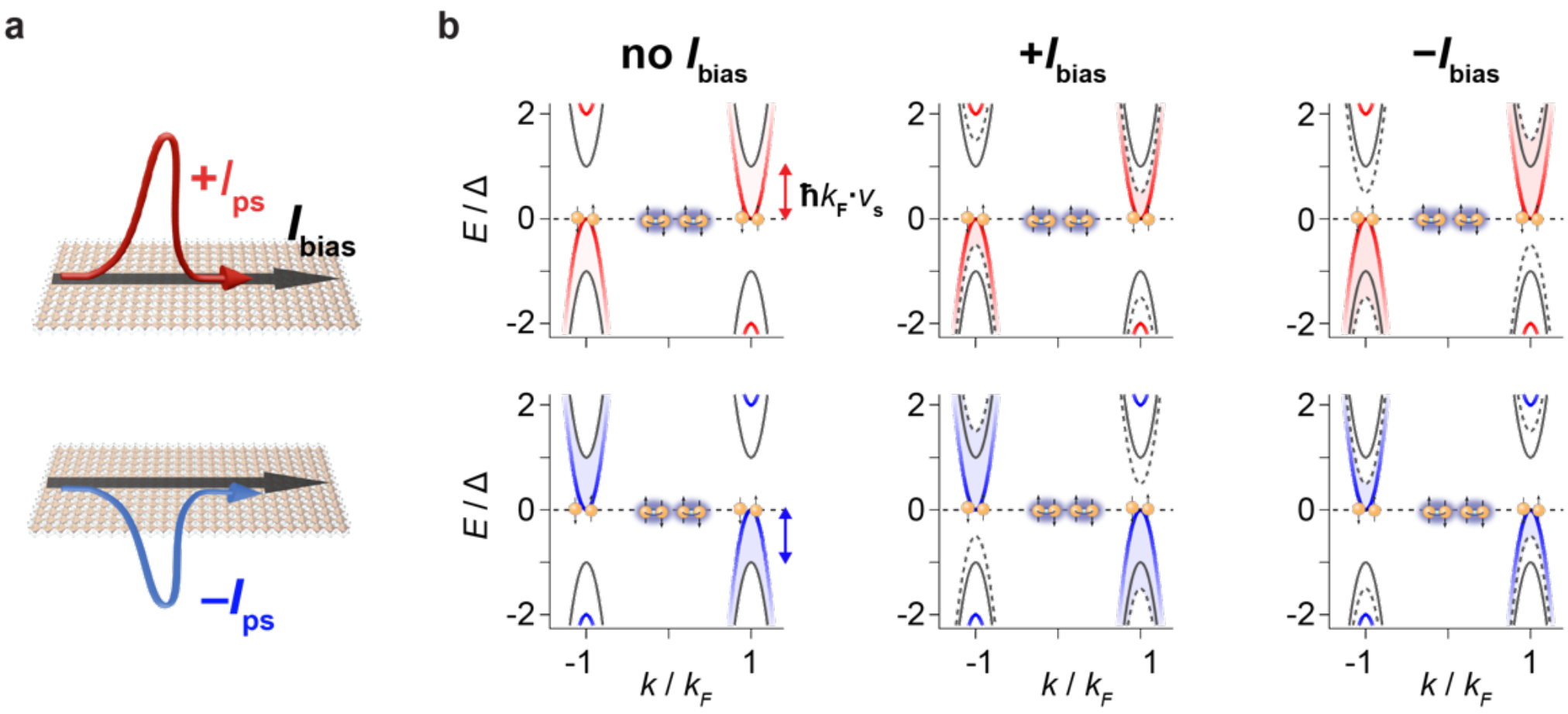

**Figure S9| Bogoliubov quasiparticle spectrum shifts induced by $\pm I_{bias}$ and $\pm I_{ps}$.** (a) Measurement schematic. (b) Schematic illustration of quasiparticle spectrum shifts under the combined application of $\pm I_{bias}$ and $\pm I_{ps}$ (top: $+I_{ps}$; bottom: $-I_{ps}$). The quasiparticle bands at zero current, after application of $\pm I_{bias}$ and after further application of $\pm I_{ps}$ are shown as solid grey, dashed grey and solid red/blue curves. The red/blue shaded regions indicate the amount of energy shift required to close the superconducting gap upon application of the picosecond current.

## S6. $I_{bias}$ amplitude dependence

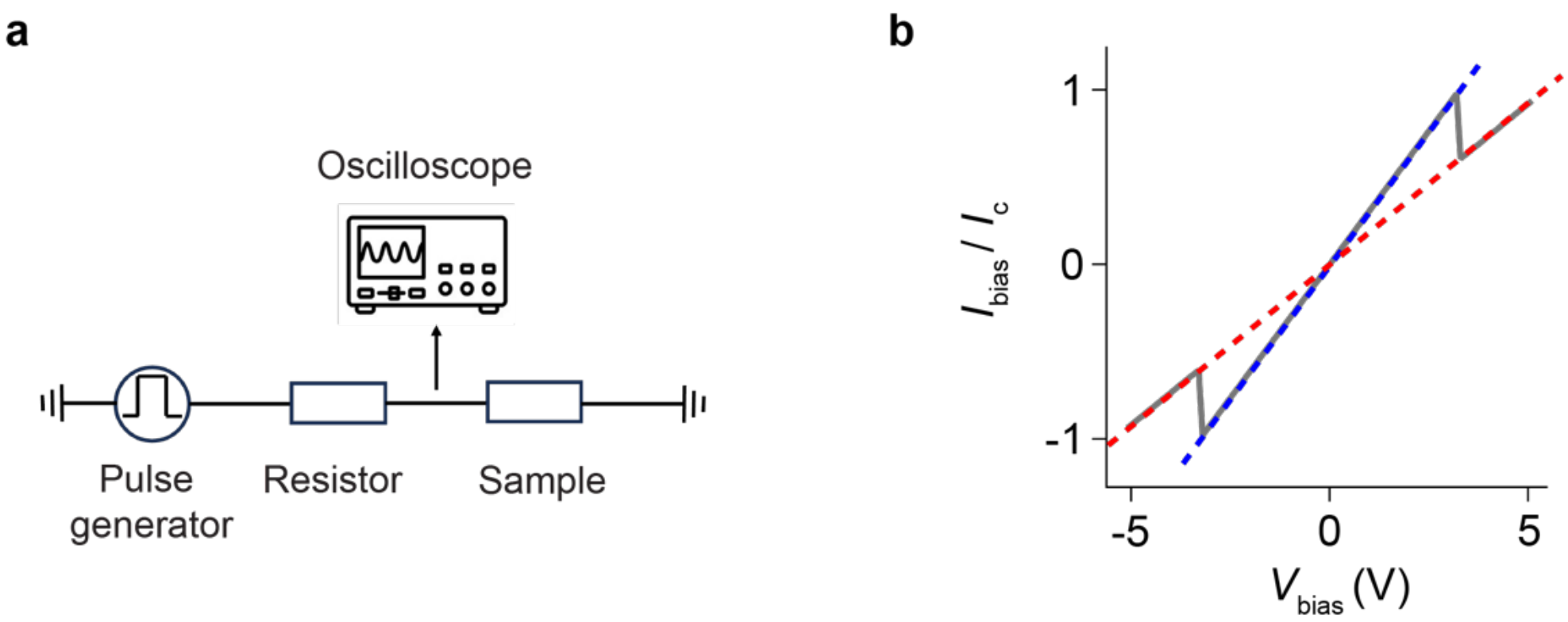

**Figure S10| Application of $I_{bias}$ with a nanosecond pulse generator.** (a) Schematic of the setup. (b) Bias current $I_{bias}$ across sample as a function of the voltage amplitude of nanosecond pulses.

To enable fast reconfiguration of the diode polarity while minimizing average Joule heating, the bias current $I_{bias}$ was applied using a nanosecond pulse generator (Agilent 81150A). The experimental schematics is shown in Fig. S10. A series pre-resistor ($R_{pre}$ = 150 Ω) was connected in series with the sample. The voltage immediately before the sample, $V_{sample}$, was monitored using an oscilloscope with an input impedance of 1MΩ. The pulse generator was configured with a load impedance $Z_{load}$ = 170 Ω and an output impedance $Z_{output}$ = 50 Ω.

The bias current flowing through the sample was calculated as

$$V_{total} = V_{bias}\frac{Z_{load}+Z_{output}}{Z_{load}} \quad \text{(S2)}$$

$$I_{bias} = \frac{V_{total}-V_{sample}}{R_{pre}+Z_{output}} \quad \text{(S3)}$$

where, $V_{bias}$ is the voltage amplitude set on the pulse generator and $V_{total}$ is the total voltage amplitude including the voltage drop across the pulse generator output impedance. The calculated bias current as a function of the output voltage is shown in Fig. S10. Two distinct linear regions are observed: the steeper slope corresponding to the superconducting state and the shallower slope corresponding to the normal state. The transition between the two regions occurs at $I_{bias} = I_c$ and is accompanied by a drop in the current.

The dependence of the asymmetric response to $\pm I_{ps}$ on the amplitudes of the picosecond current pulses and the bias current is shown in Fig. S11. Fig. S11a presents the normalized peak electric field of the transmitted pulse, *Norm*. $E_{T,peak}$ (normalized by that of incident pulse), as a function of incident peak current density $J_{in,peak}$. To avoid driving the sample into the normal state before picosecond pulses arrived, the maximum bias current amplitude was limited to $I_c$. As a result, the asymmetric response between $+I_{ps}$ and $-I_{ps}$ is maximized when the incident current density lies near the midpoint of the slope in Fig. S11a, as shown in Fig. S11b-g.

In all cases, *Norm*. $E_{T,peak}$ drops to the normal-state value when the nanosecond bias pulse drives the sample into the normal state ($|V_{bias}| > 3.5$ V, where $I_{bias}$ abruptly drops shown in Fig. S10b). The slight downturn of the *Norm*. $E_{T,peak}$ observed near $|V_{bias}| \approx 3.5$ V is most likely associated with the kinetic response of the sample due to the progressive suppression of superconductivity by the picosecond current pulse.

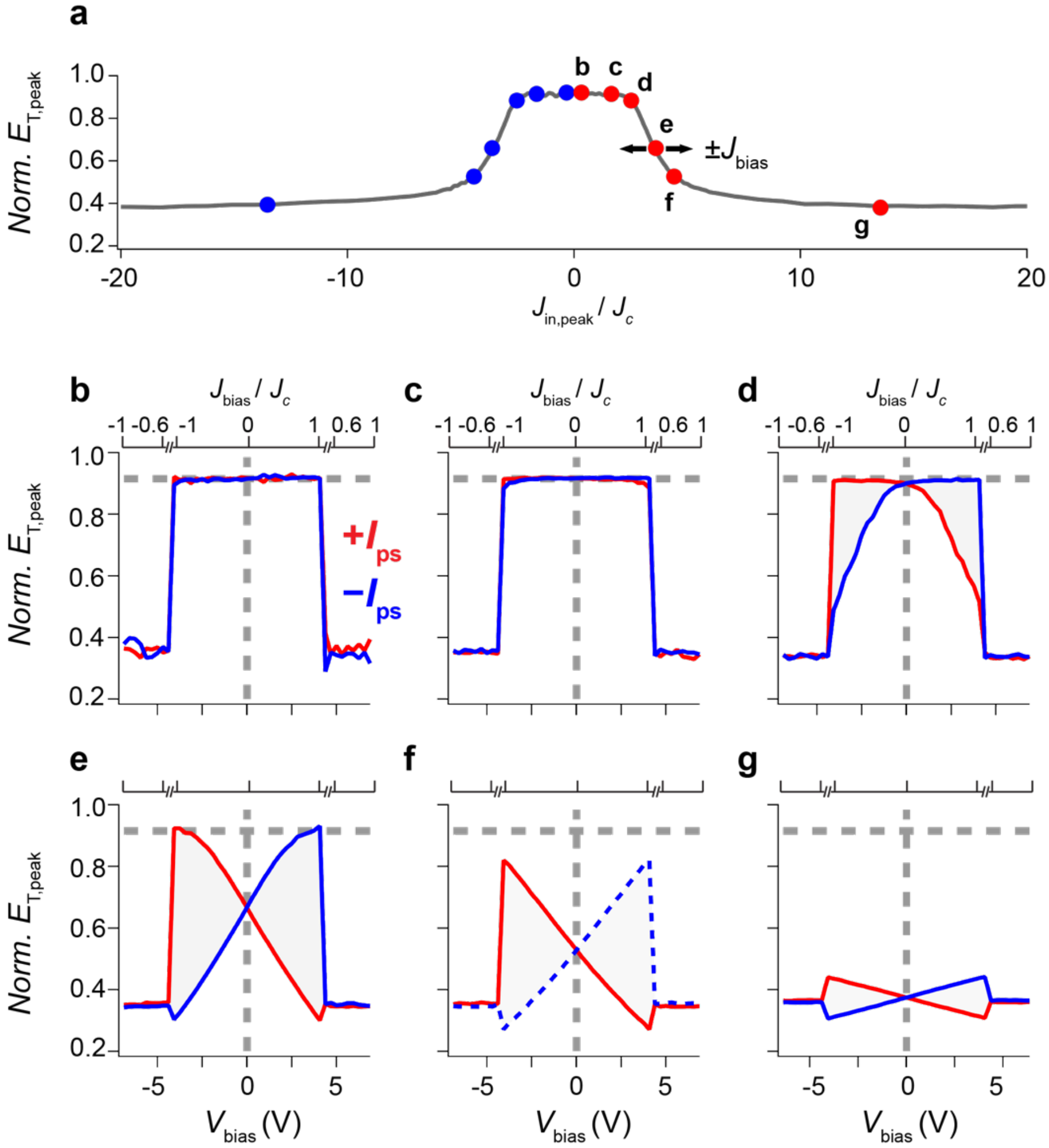

**Figure S11| Dependence of the diode effect on the amplitudes of the picosecond current pulses and the bias current.** (a) Normalized peak electric field of the transmitted pulse, *Norm.* $E_{T,peak}$ (normalized by that of the incident pulse), as a function of the incident peak current density $J_{in,peak}$. (b-g) Change in *Norm.* $E_{T,peak}$ induced by the application of $I_{bias}$ for picosecond pulses with different amplitudes. The horizontal line indicated the superconducting-state value. The dashed blue curve in panel (f) is obtained by direct horizontal inversion of the red curve.

## S7. Switch diode polarity on nanosecond time scale

The polarity of the diode effect can be reversed by switching the direction of the bias current. Using a nanosecond pulse generator, this polarity switching can be achieved on nanosecond timescales. In this measurement, the end of the sample was connected to an oscilloscope with an input impedance 50 Ω, rather than directly grounded, to both monitor the current flowing through the sample and ensure proper impedance matching. As shown in Fig. 12(b,c), reversing the bias current switches the diode polarity within approximately 10ns.

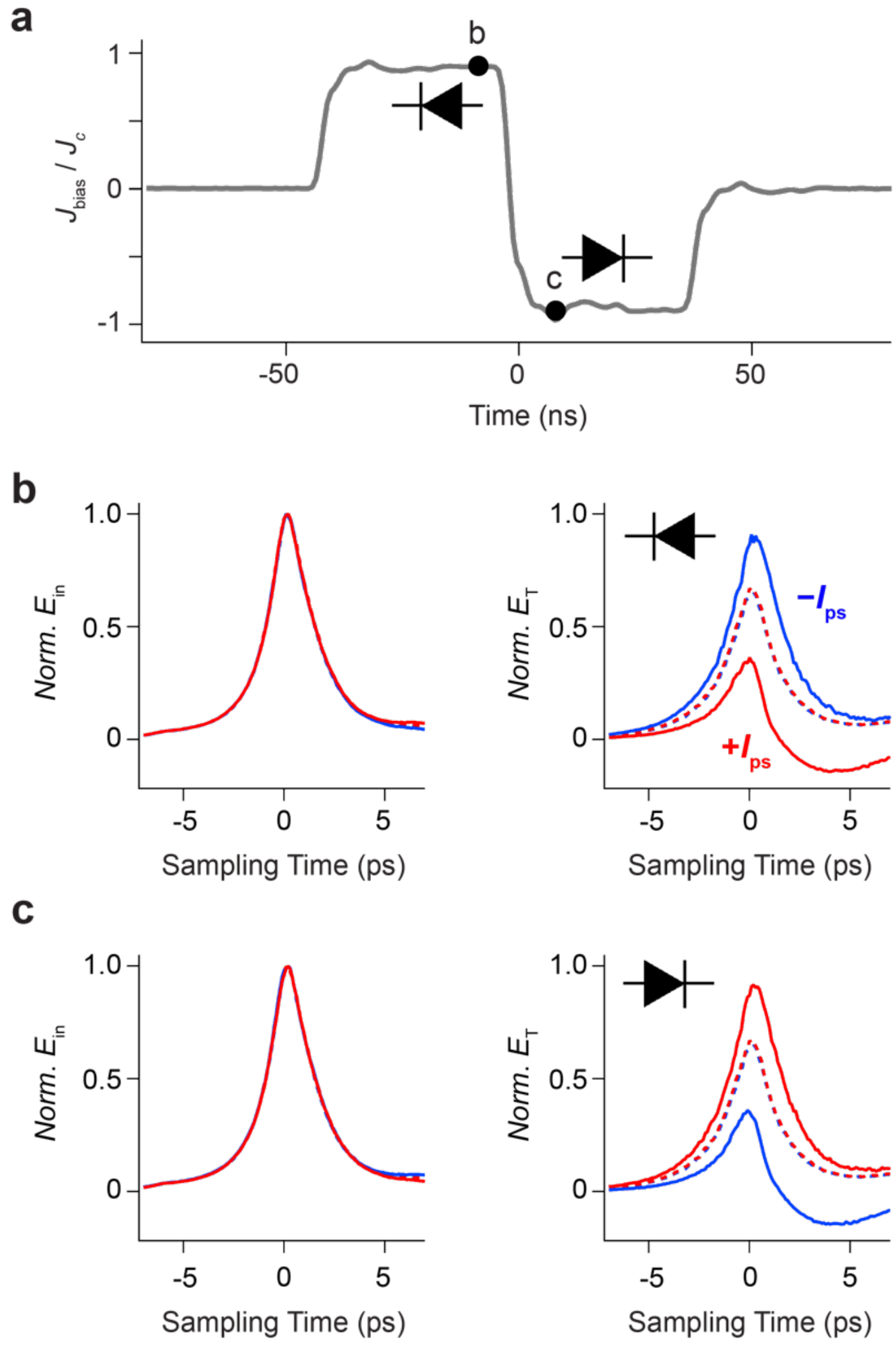

**Figure S12| Diode polarity switching.** (a) Time profile of the bias current flowing through the sample. (b-c) Time profiles of the normalized electric field of the incident and transmitted pulses (normalized by the peak field of the incident pulse) under positive (b) and negative (c) bias current. The dashed curves in panels (b) and (c) show the corresponding responses measured at zero bias current. In all measurements, the peak current density of the incident picosecond pulse was 3.2 $J_c$.

## S8. Generation of bipolar pulses train

The method used to generate the bipolar pulse train in this study is illustrated in Fig. S13. A single 515-nm laser beam was split twice with two 50/50 beamsplitters and subsequently guided into the cryostat. Four delay stages were used to adjust the relative timing of the four laser pulses, producing an even temporal separation of 5 ps. The two pairs of double laser pulses were directed onto two separate photoconductive switches, biased at $+V$ and $-V$, respectively. The incident and the transmitted picosecond current pulses were detected using photoconductive switches located to the left of the launch point and on the right side of the sample, respectively.

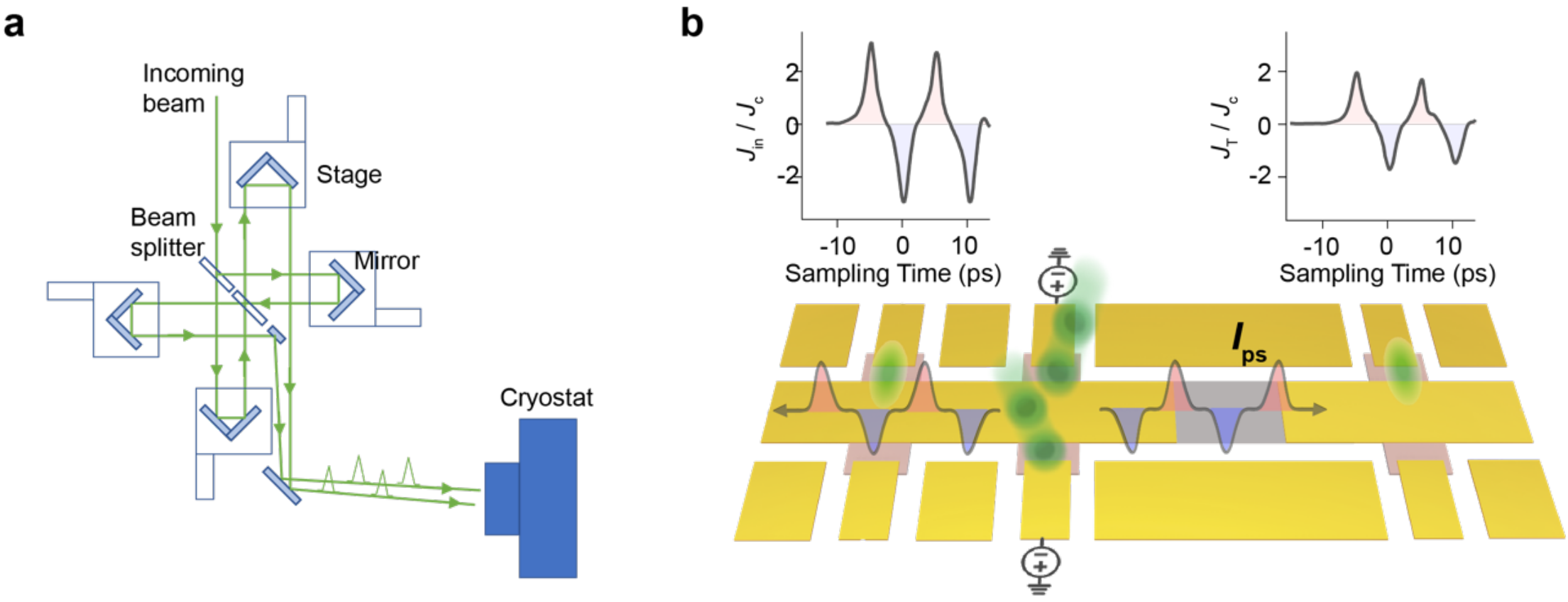

**Figure S13| Sketch of bipolar pulse train generation.** (a) Generation of two sets of double laser pulses, with all pulses evenly separated by 5 ps. (b) Detection of the incident and transmitted bipolar pulse trains.

## S9. Estimation of energy dissipation

The energy dissipation is calculated by taking the energy difference between the incoming pulse and the sum of reflected plus transmitted pulses. The energy contained in each pulse is calculated by summing up the energy stored in the E- and H-field.

The energy stored in the E-field is:

$$E_E = \int \frac{1}{2} C_0 V^2 dx = \int \frac{1}{2} C_0 (Z_0 I)^2 v dt$$

where $C_0 \sim 0.16\, fF$ is the capacitance of coplanar waveguide per $\mu m$, $v \sim 130\ \mu m/ps$ is the velocity of the pulse propagating along the waveguide and $Z_0 \sim 50\Omega$ is the impedance of the coplanar waveguide.

Similarly, the energy stored in the H-field is

$$E_H = \int \frac{1}{2} L_0 I^2 dx = \int \frac{1}{2} L_0 I^2 v dt$$

Where $L_0 \sim 0.37\, pH$ is the inductance of the coplanar waveguide per $\mu m$.

The time profiles of incoming, transmitted, and reflected pulses are taken from data shown in Ref. [2]. For 8-µm-wide, 30-µm-long, and 20-nm-thick NbN strip used in this paper, operating at a peak current ~ 3 $J_c$ = 45 mA, where the diode effect is strongest, the dissipations are 3 fJ and 53 fJ when sample is superconducting and resistive, respectively. With straightforward device miniaturization to reduce sample dimensions to 1-µm-wide, 1-µm-long and 10-nm-thick, the correspond energy dissipations are reduced to 6 and 106 aJ correspondingly.

## S10. Simulations of asymmetric picosecond responses

The simulations of depairing dynamics in a NbN strip within the BCS framework in the dirty limit, are discussed in a previous paper, see Ref. [2].

To account for the effect of the DC bias, we repeat the same protocol as before, however changing the initial state of the system to correspond to the equilibrium superconducting state with a DC bias current corresponding to $I = \pm 0.97\ I_c$ (±19.5 mA). In our simulations, we choose the value for Dynes broadening $\eta$ equal to 15% of the $T_c^{BCS}$, to ensure numerical stability and convergence. The finite value of $\eta$ causes corrections in the gap temperature dependence and critical temperature. Therefore, in the fitting procedure, we choose the BCS coupling to correspond to $T_c(\eta = 0) = T_c^{BCS} = 1$ and represent all the quantities in units of $T_c^{BCS}$. For the chosen value of Dynes broadening, we obtain $T_c = 0.84\, T_c^{BCS}$, therefore$T_c^{BCS} = \frac{1}{0.84} * 14.5\ K$. All other parameters are extracted and fitted using the same way as in the previous paper [2].